\documentstyle[epsf]{aipproc}

\begin{document}

\title{Black Holes with Yang-Mills Hair}
\author{B. Kleihaus, J. Kunz, A. Sood and M. Wirschins\thanks{supported
by DFG}}
\address{Fachbereich Physik, Universit\"at Oldenburg, Postfach 2503\\
D-26111 Oldenburg, Germany}
\maketitle

\begin{abstract}

In Einstein-Maxwell theory
black holes are uniquely determined by their
mass, their charge and their angular momentum. 
This is no longer true in Einstein-Yang-Mills theory.
We discuss sequences of neutral and charged $SU(N)$
Einstein-Yang-Mills black holes, which are
static spherically symmetric and asymptotically flat,
and which carry Yang-Mills hair.
Furthermore, in Einstein-Maxwell theory 
static black holes are spherically symmetric.
We demonstrate that, in contrast, $SU(2)$ Einstein-Yang-Mills theory
possesses a sequence of black holes,
which are static and only axially symmetric.

\end{abstract}

\section{Introduction}

The ``no hair'' conjecture for black holes states,
that black holes are uniquely characterized 
by their mass, their charge and their angular momentum.
This conjecture presents a generalization of
rigorous results obtained for scalar fields coupled to gravity
as well as for Einstein-Maxwell (EM) theory.
Notably, the static black hole solutions in EM theory
are spherically symmetric, and the stationary black holes
are axially symmetric.

In recent years, counterexamples to the ``no hair'' conjecture
were established in various theories with non-abelian fields,
including Einstein-Yang-Mills (EYM) theory,
Einstein-Yang-Mills-dilaton (EYMD) theory,
Einstein-Yang-Mills-Higgs (EYMH) theory,
and Einstein-Skyrme (ES) theory.
These non-abelian black holes are asymptotically flat
and possess a regular event horizon with
non-trivial matter fields outside the event horizon.

In $SU(2)$ EYM theory,
there exists a sequence of neutral static spherically symmetric
non-abelian black hole solutions, labelled by
the node number $n$ of the gauge field function \cite{eym}.
Existing for arbitrary horizon radius $x_{\rm H}$,
these black hole solutions with Yang-Mills hair
approach globally regular solutions
\cite{bm} in the limit $x_{\rm H} \rightarrow 0$.
In contrast, all static spherically symmetric $SU(2)$
EYM black hole solutions with non-zero charge are embedded
Reissner-Nordstr\o m (RN) solutions \cite{ersh}.
However, this ``non-abelian baldness theorem'' no longer holds
for $SU(3)$ EYM theory \cite{gv}.

Non-abelian static spherically symmetric solutions
of $SU(N)$ EYM theory are obtained by
embedding the $N$-dimensional representation of $su(2)$ in $su(N)$.
The gauge field ansatz then involves
$N-1$ gauge field functions, 
and the solutions are labelled
by the corresponding node numbers $(n_1,\ldots,n_{N-1})$
\cite{kuenzle,kks2,kksw,sun}.
When all gauge field functions are non-trivial, 
neutral globally regular and black hole solutions are obtained.
But when one or more of these functions are identically zero, 
magnetically charged black hole solutions emerge,
whose charge resides in the $su(N)$ Cartan subalgebra (CSA).
All these solutions possess EYMD counterparts \cite{eymd,kks2,kksw,sun}.

We here first consider static spherically symmetric
$SU(N)$ EYM and EYMD solutions,
presenting examples of neutral globally regular $SU(4)$ solutions
and charged $SU(5)$ black hole solutions,
and consider also the interior of the black hole solutions.

Then we discuss globally regular and black hole solutions 
of $SU(2)$ EYM and EYMD theory, which are asymptotically flat, static
and only axially symmetric but not spherically symmetric \cite{kk3}.
These solutions are characterized by two integers,
the winding number $k>1$
and the node number $n$ of the gauge field functions.                          
The black hole solutions possess a regular event horizon
and a constant surface gravity.
We argue, that non-abelian theories even possess
static black hole solutions with only discrete symmetries.

\boldmath
\section{$SU(N)$ EYMD action}
\unboldmath

We here consider the action of $SU(N)$ EYMD theory
\begin{equation}
S=\int \left ( \frac{R}{16\pi G} + L_M \right ) \sqrt{-g} d^4x
\   \end{equation}
with matter Lagrangian
\begin{equation}
L_M=-\frac{1}{2}\partial_\mu \Phi \partial^\mu \Phi
 -e^{2 \kappa \Phi }\frac{1}{2} {\rm Tr} (F_{\mu\nu} F^{\mu\nu})
\ , \end{equation}
field strength tensor
$F_{\mu \nu} = 
\partial_\mu A_\nu -\partial_\nu A_\mu + i e \left[A_\mu , A_\nu \right] $,
gauge field
$ A_\mu = \frac{1}{2} \lambda^a A_\mu^a$,
dilaton field $\Phi$,
and $e$ and $\kappa$ are the Yang-Mills and dilaton coupling constants,
respectively.

\section{Static spherical black holes}

As shown by Bartnik and McKinnon \cite{bm}, $SU(2)$ EYM theory possesses
a sequence of neutral globally regular solutions. These have black hole
counterparts with a regular horizon \cite{eym}. 
$SU(N)$ EYM theory possesses a with $N$ increasing number of
sequences of neutral globally regular and black hole solutions 
\cite{kks2,kksw}
as well as of charged black hole solutions \cite{gv,sun},
which all have EYMD counterparts.

\subsection{Ans\"atze}

To construct static spherically symmetric $SU(N)$ EYM and EYMD solutions
we employ Schwarz\-schild-like coordinates and adopt
the spherically symmetric metric
\begin{equation}
ds^2=g_{\mu\nu}dx^\mu dx^\nu=
  -{\cal A}^2{\cal N} dt^2 + {\cal N}^{-1} dr^2 
  + r^2 (d\theta^2 + \sin^2\theta d\phi^2)
\ , \label{metric} \end{equation}
with the metric functions ${\cal A}(r)$ and 
${\cal N}(r)=1-({2m(r)}/{r})$.

The static spherically symmetric ans\"atze
for the gauge field of $SU(N)$ EYM theory
are based on the $su(2)$ subalgebras of $su(N)$.
Considering the $N$-dimensional representation of $su(2)$,
the ansatz is \cite{kuenzle}
\begin{equation} \label{ansa}
 A_{\mu}^{(N)}  dx^\mu   =  \frac{1}{2e} \left( 
\begin{array}{ccccc}
(N-1)\cos\theta d\phi & \omega_1 \Theta & 0 & \ldots & 0 \\
\omega_1 \bar \Theta & (N-3)\cos\theta d\phi & \omega_2 \Theta & 
\ldots & 0 \\
\vdots & & \ddots & & \vdots \\
0 & \ldots & 0 & \omega_{N-1} \bar \Theta & (1-N)\cos\theta d\phi
\end{array} \right)
\label{amu} \end{equation}   
with $\Theta = i d \theta + \sin \theta d \phi$,
and $A_0=A_r=0$.
The ansatz contains $N-1$ matter field functions $\omega_j(r)$.
The field strength tensor component
$F_{\theta\phi}$ is diagonal,
\begin{equation} \label{F1}
F_{\theta\phi}=(1/2e) {\rm diag} (f_1,...,f_N) \sin \theta
\   \end{equation}   
with
$ f_j = \omega_j^2 - \omega^2_{j-1} + \delta_j$,
$\delta_j = 2j - N -1$ ($\omega_0=\omega_N=0$).
In EYMD theory this is supplemented with the ansatz 
for the dilaton field, $\Phi=\Phi(r)$.

We employ the dimensionless coordinate 
$x=er/{\sqrt{4\pi G}}$,
the dimensionless mass function
$\mu=em/{\sqrt{4\pi G}}$,
and the scaled matter field functions \cite{kuenzle}
\begin{equation}
 u_j = \frac{\omega_j}{\sqrt{\gamma_j}} \ , \ \ \
 \gamma_j = {j (N - j) } 
\  , \end{equation}
and, in EYMD theory,
the dimensionless dilaton function $\varphi = \sqrt{4\pi G} \Phi$ and
the dimensionless dilaton coupling constant
$\gamma =\kappa/\sqrt{4\pi G}$.
($\gamma=1$ corresponds to string theory,
 $\gamma=0$ to EYM theory.)

\subsection{Neutral solutions}

When all $N-1$ gauge field functions are non-trivial,
neutral solutions are obtained.
The boundary conditions are chosen to have
asymptotically flat solutions
with a regular origin for the globally regular solutions
and a regular horizon for the black hole solutions.

As an example, we consider static spherically symmetric 
globally regular solutions of $SU(4)$ EYM theory \cite{kksw}.
The $SU(4)$ EYM solutions
are labelled by the node numbers $(n_1,n_2,n_3)$ of the three
gauge field functions $u_1$, $u_2$ and $u_3$,
and can be classified into sequences.
In Fig.~1 we present the globally regular solutions 
of the sequence with node structure $(n,0,0)$, $n=1-7$, with odd $n$.
With increasing $n$, the solutions approach a limiting solution.
\begin{figure}[t]
\vspace*{0.5cm}
\centering
 \hspace*{-4.5cm} 
\makebox[11.cm][r]{\epsfysize=5.cm \epsffile{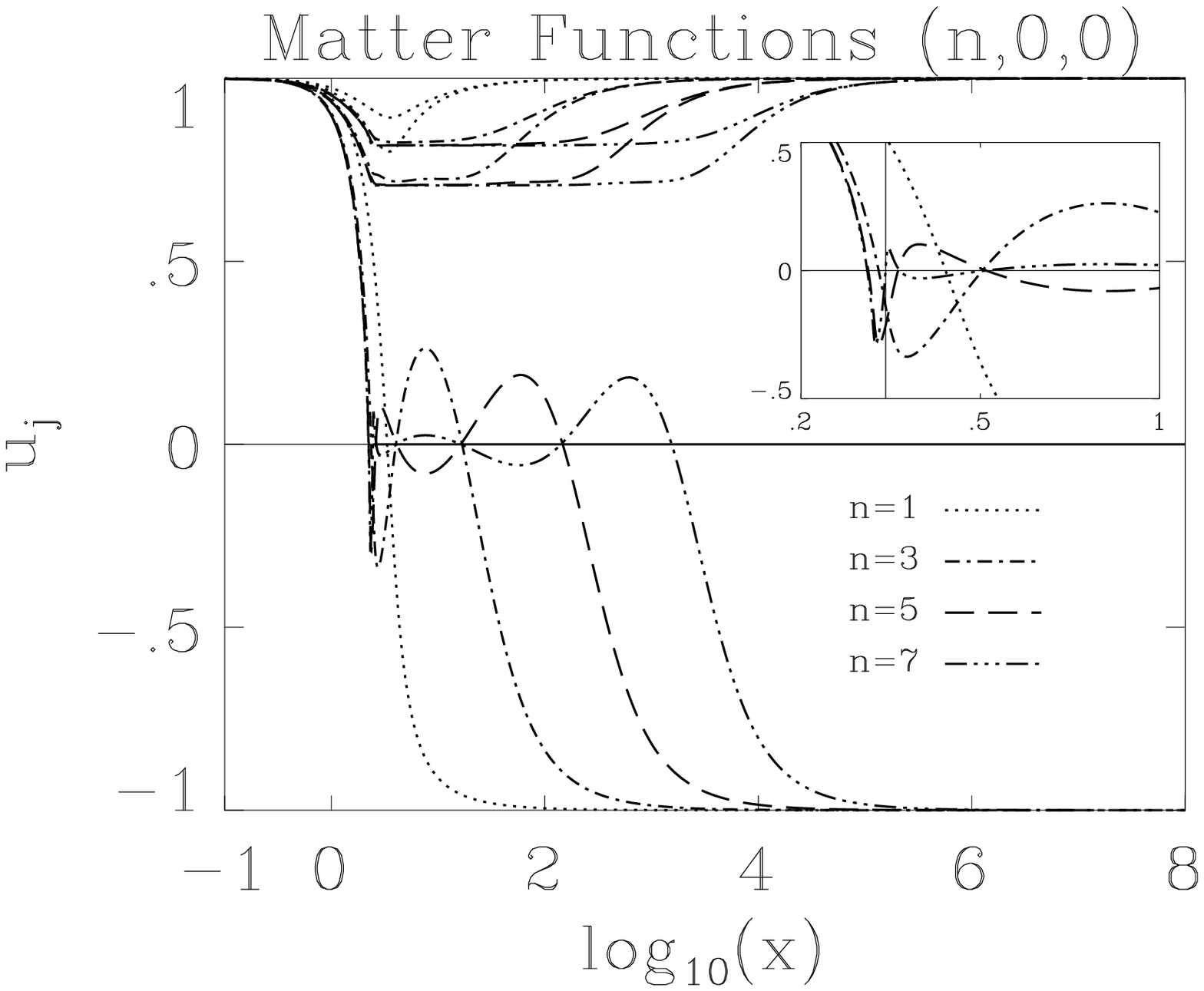} }
 \hspace*{-4.0cm} 
\makebox[11.cm][r]{\epsfysize=5.cm \epsffile{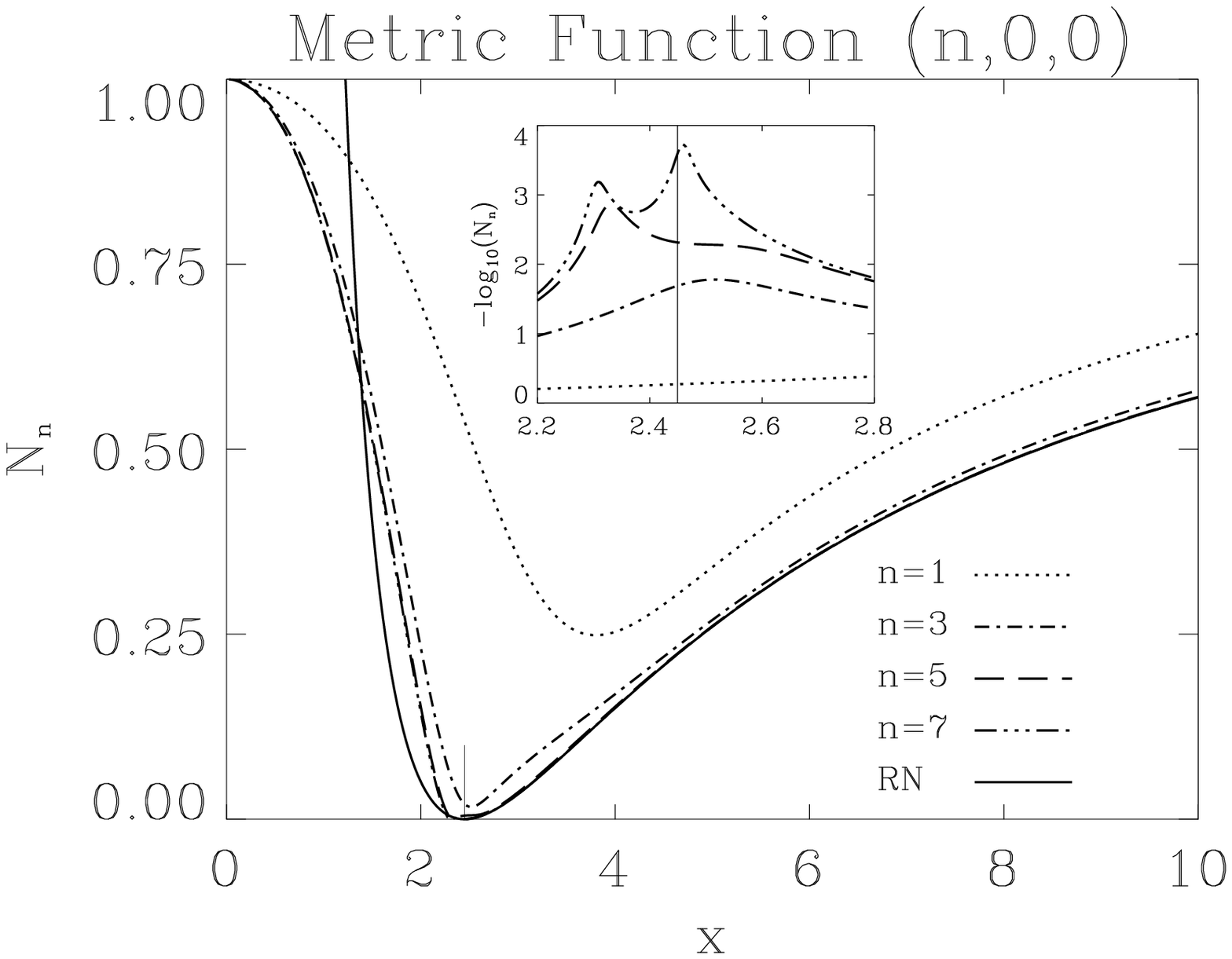} }
\caption{
The matter functions $u_{1}(x)$, $u_{2}(x)$ and $u_{3}(x)$ (left)
and the metric function ${\cal N}(x)$ (right)
for the globally regular $SU(4)$ EYM solutions
with node structure $(n,0,0)$.
}
\end{figure}

To all regular solutions black hole counterparts exist.
When all node numbers differ, these black hole solutions exist
for arbitrary horizon radius. When two or more node numbers are the
same, bifurcation phenomena occur at critical horizon radii
\cite{kks2,kksw}.

\subsection{Charged solutions}

When one or more of the $N-1$ gauge field functions 
are identically zero,
magnetically charged solutions are obtained.

\subsubsection{Decomposition of the ansatz}

When one gauge field function is identically zero
($\omega_{j_{1}} \equiv 0$),
the ansatz for the gauge field splits into two parts
\begin{equation}
\begin{array}{ccc}
A_\mu^{(N)} dx^\mu & = & 
\left( 
\begin{array}{cc}
{\rm \fbox{$ A_\mu^{(j_1)} dx^\mu  $}} &                      \\
                     &  {\rm \fbox{$A_\mu^{(N-j_1)} dx^\mu $}}\\
\end{array}
\right)
 + {\cal H}_{j_1}
\end{array} \vspace{1.cm} 
\ \label{amu1} \end{equation}
with  
${\cal H}_{j_1}  =  \frac{\cos\theta d\phi}{2 e} h_{j_1}$ and
\begin{equation}
\begin{array}{ccc}
{h}_{j_1} & = & 
\left( 
\begin{array}{cc}
{\rm \fbox{$(N-j_1){\bf 1}_{(j_1)} $}}&     \\
      &{\rm \fbox{$ -j_1 {\bf 1}_{(N-j_1)}$}}   \\            
\end{array}
\right)\\
\end{array}\vspace{1.cm} 
\ . \label{h1} \end{equation}
$A_\mu^{(j_1)}$ and $A_\mu^{(N-j_1)}$ denote the non-abelian
spherically symmetric ans\"atze for the $su(j_1)$
and $su(N-j_1)$ subalgebras of $su(N)$
(based on the $j_1$ and $(N-j_1)$-dimensional embeddings of $su(2)$,
respectively), referred to by $su(\bar{N})$ in the following.
${\cal H}_{j_1}$ represents the ansatz for the element 
${h}_{j_1}$ of the CSA of $su(N)$.
The field strength tensor splits accordingly with
\begin{equation}
F^{({\cal H}_{j_1})} = -\frac{\sin \theta}{2e} d\theta \wedge d\phi \ h_{j_1}
\ . \label{FH}\end{equation}

Considering the charge of the solution, 
the $su(\bar N)$ parts of the solutions are neutral, because
their field strength tensor decays at least like $O(r^{-1})$.
In contrast, $F^{({\cal H}_{j_1})}$ does not depend on $r$.
A solution based on the element $h_{j_1}$ of the CSA then carries
charge of norm $P$,
\begin{equation}  
P^2 = \frac{1}{2}{\rm Tr} \ {h}_{j_1}^2
\ . \label{Pa} \end{equation}
By applying these considerations again to the subalgebras
$su(\bar N)$ of eq.~(\ref{amu1}), one obtains
the general case for $SU(N)$ EYM theory \cite{sun}.

RN black hole solutions exist only for horizon radius
$x_{\rm H}\ge P$, and
the extremal RN solution has $x_{\rm H}=P$.
As first observed for $SU(3)$ EYM theory \cite{gv},
the same is true for charged EYM black holes.
In contrast, charged EYMD black holes exist for arbitrary
$x_{\rm H} > 0$, like their Einstein-Maxwell-dilaton (EMD)
counterparts.

\subsubsection{Extremal black hole solutions}

We now apply the above general analysis
to $SU(5)$ EYM theory, presenting numerical examples
for the case $u_4\equiv 0$ \cite{sun}.
In that case the non-abelian part of the gauge field corresponds to
an $su(4)$ part, and the solutions carry charge $P=\sqrt{10}$.
In Fig.~2 and Fig.~3 (left) we present the black hole solutions 
of the sequence with node structure $(n,0,0)$, $n=1-7$, with odd $n$,
and extremal horizon $x_{\rm H}=\sqrt{10}$,
also in the interior of the black hole (for $x<x_{\rm H}$).
Beside the matter function $\bar u_1$ we exhibit
the charge function $P(x)$,
$P^2(x)= 2 x \left(\mu(\infty)-\mu(x) \right) $ \cite{bm,gv},
and the metric function ${\cal N}(x)$.
\begin{figure}[h]
\vspace*{0.5cm}
\centering
 \hspace*{-4.5cm} 
\makebox[11.cm][r]{\epsfysize=5.cm \epsffile{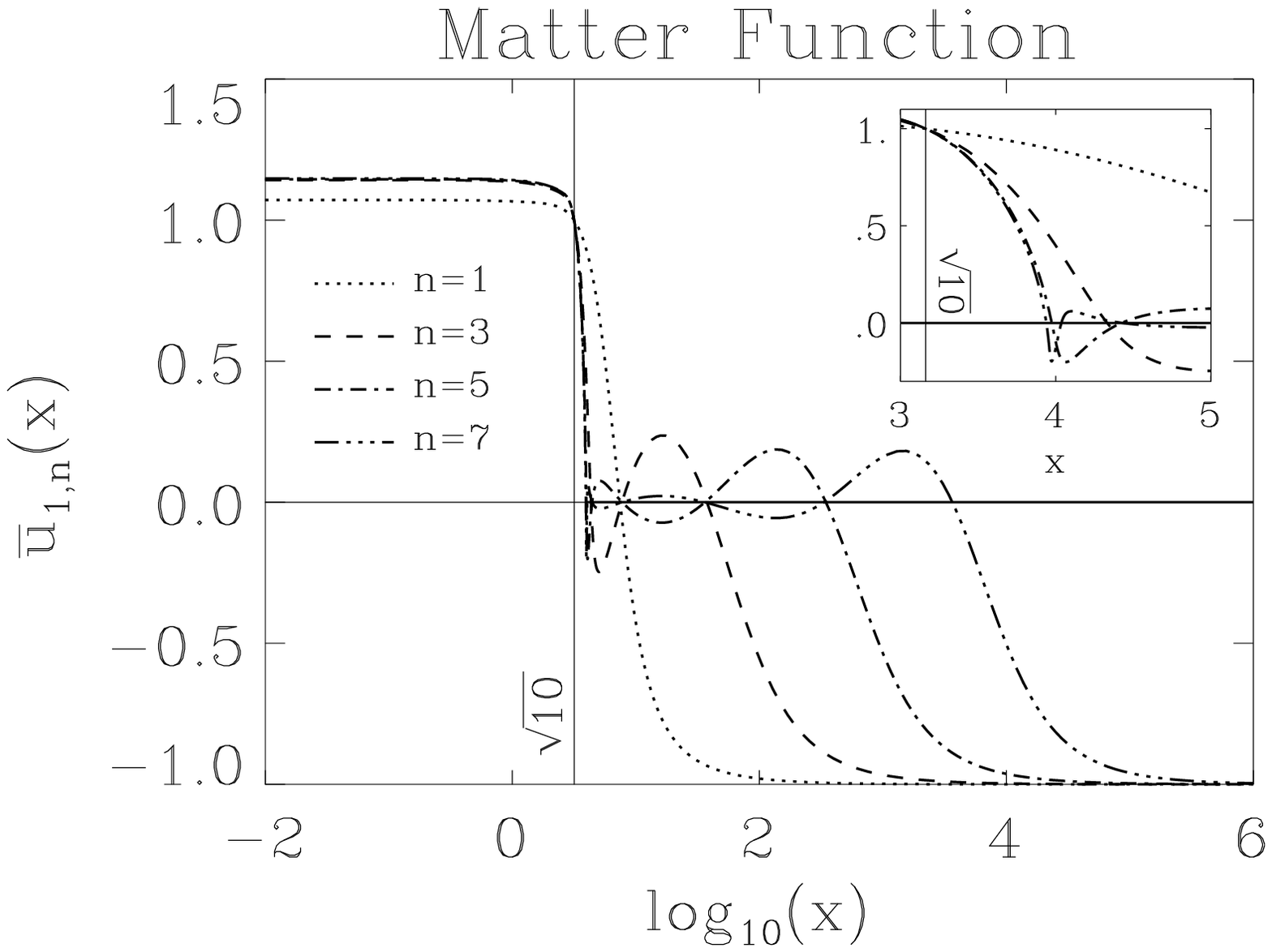} }
 \hspace*{-4.0cm} 
\makebox[11.cm][r]{\epsfysize=5.cm \epsffile{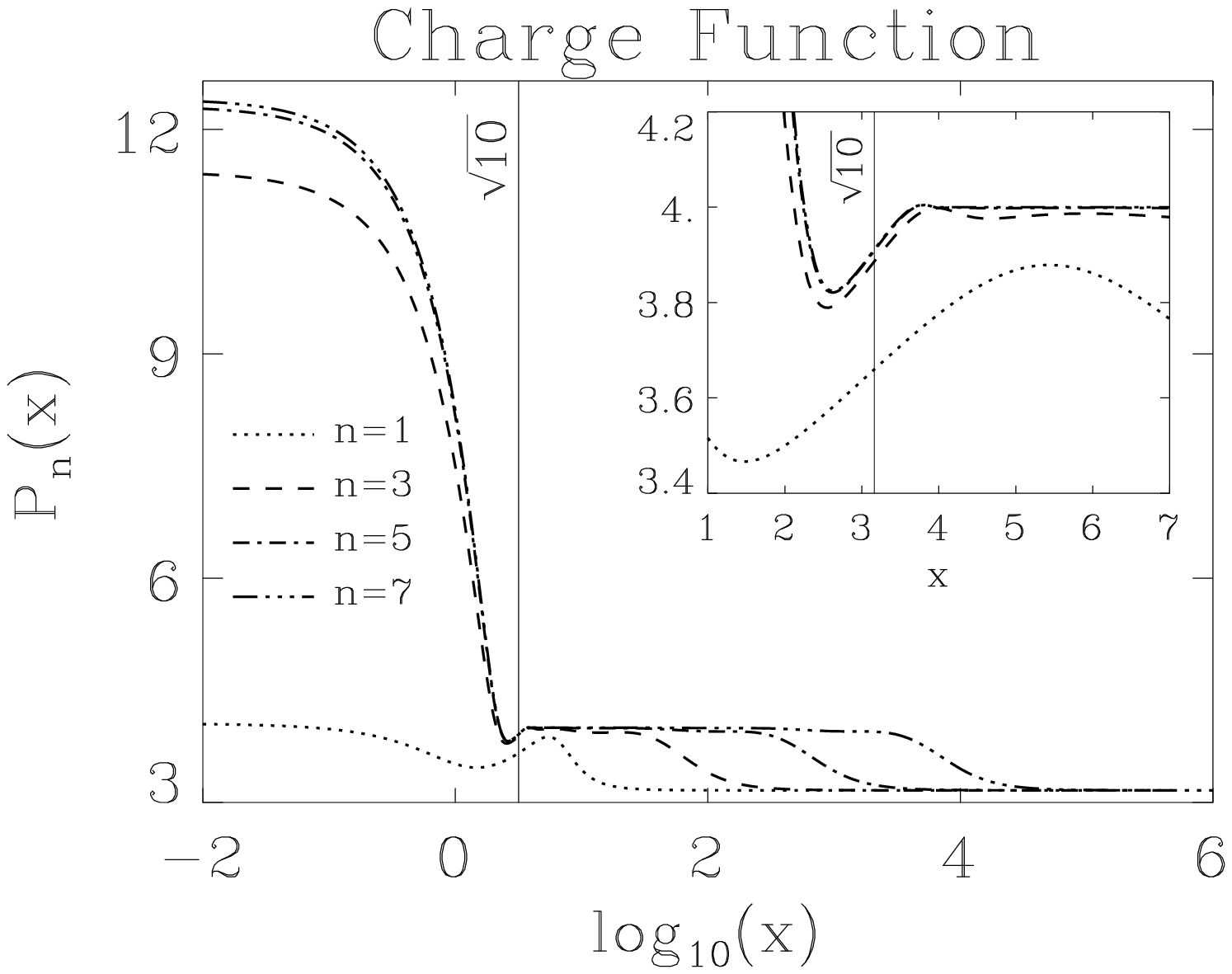} }
\caption{
The matter function $\bar u_{1}(x)$ (left)
and the charge function $P(x)$ (right)
for the $SU(4)$ EYM black hole solutions
with node structure $(n,0,0)$
and extremal event horizon $x_{\rm H} = \protect\sqrt{10}$.
}
\end{figure}
\begin{figure}[h!]
\vspace*{0.5cm}
\centering
 \hspace*{-4.5cm} 
\makebox[11.cm][r]{\epsfysize=5.cm \epsffile{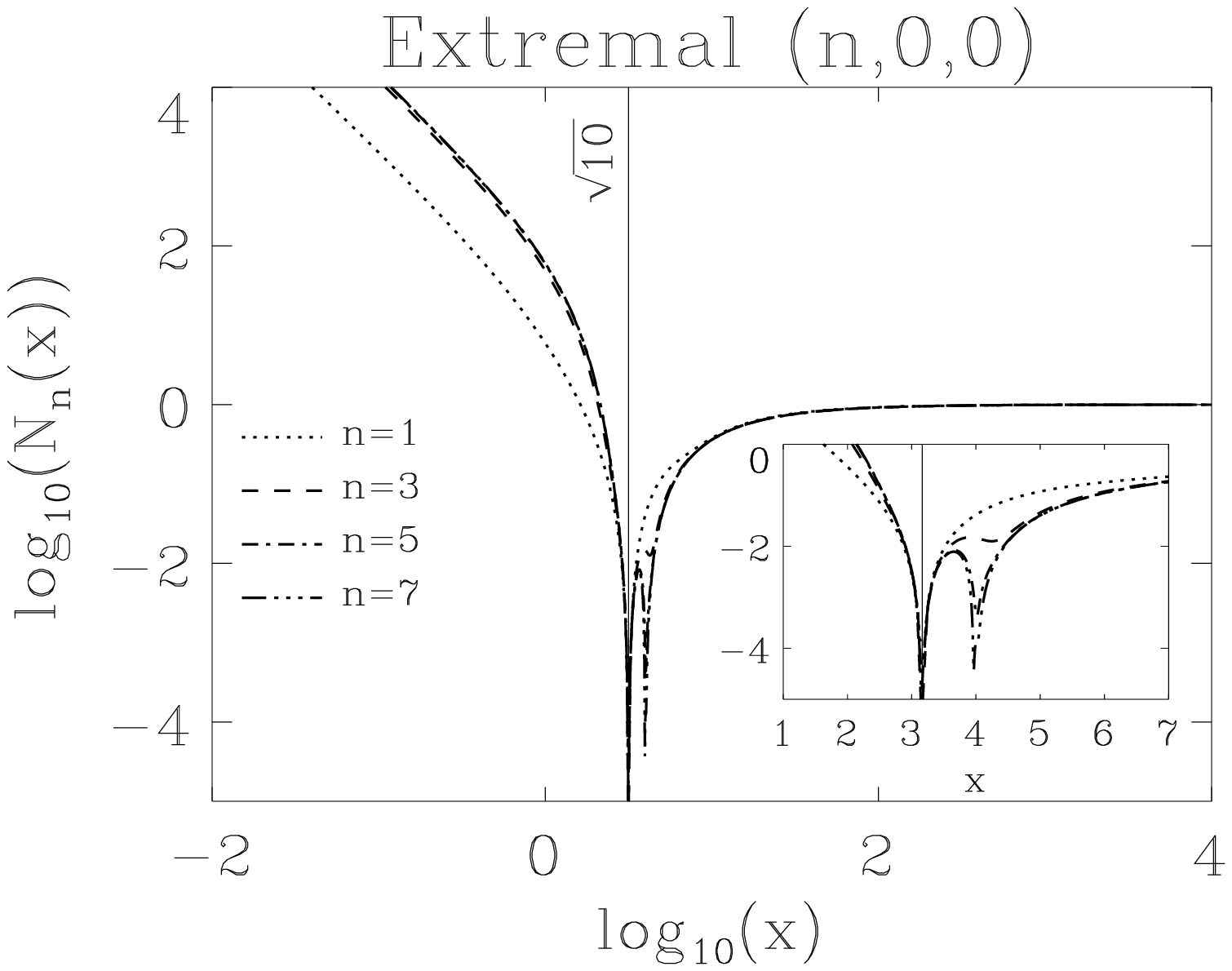} }
 \hspace*{-4.0cm} 
\makebox[11.cm][r]{\epsfysize=5.cm \epsffile{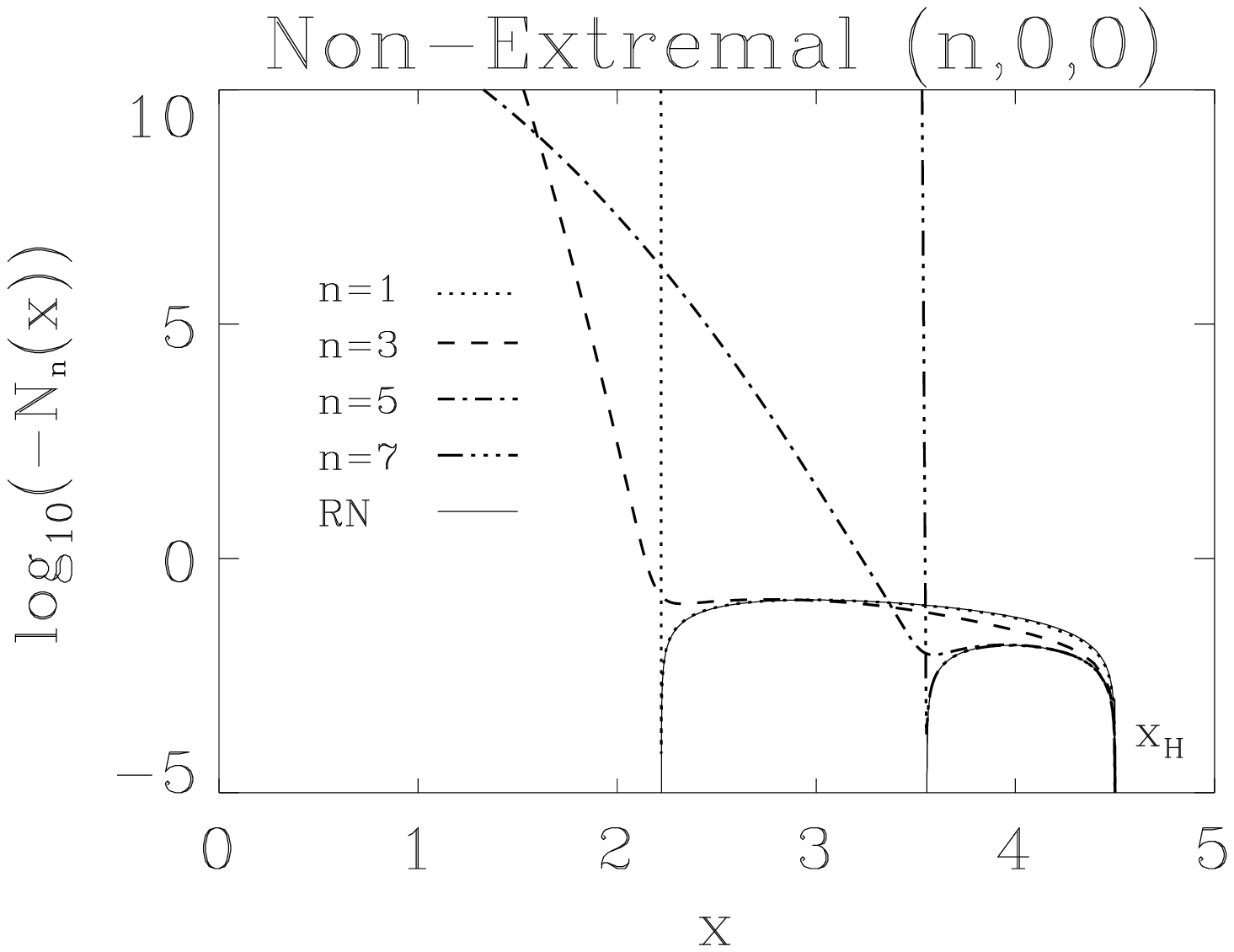} }
\caption{
The metric function ${\cal N}(x)$
for the $SU(4)$ EYM black hole solutions
with node structure $(n,0,0)$
and extremal event horizon $x_{\rm H} = \protect\sqrt{10}$ (left)
and non-extremal event horizon $x_{\rm H} = 4.5$ (right).
Also shown (right) are the corresponding functions of the RN solutions
with charges $P=\protect\sqrt{10}$ and $P=4$.
}
\end{figure}

\subsubsection{Non-extremal black hole solutions}

As seen from  Fig.~2 and Fig.~3 (left), extremal EYM black hole solutions 
vary smoothly inside the horizon.
In contrast, non-extremal EYM black hole solutions exhibit
the phenomenon of mass inflation inside the horizon, 
analogously to neutral EYM black hole solutions \cite{massi}.
In Fig.~3 (right) we demonstrate this phenomenon for the metric function
${\cal N}$ for the same sequence with node structure $(n,0,0)$, $n=1-7$, 
$n$ odd,
and non-extremal horizon $x_{\rm H}=4.5$.
In contrast, for EYMD solutions with $\gamma=1$ no mass inflation occurs.

\section{Static axial black holes}

\subsection{Ans\"atze}

To obtain static axially symmetric solutions,
we now employ isotropic coordinates for the metric \cite{kk3,kk2}
\begin{equation}
ds^2=
  - f dt^2 +  \frac{m}{f} d r^2 + \frac{m r^2}{f} d \theta^2 
           +  \frac{l r^2 \sin^2 \theta}{f} d\phi^2
\ , \label{ametric} \end{equation}
where $f$, $m$ and $l$ are only functions of $r$ and $\theta$.
We parametrize the purely magnetic gauge field ($A_0=0$) 
by \cite{kk3,kk2,rr}
\begin{equation}
A_\mu dx^\mu =
\frac{1}{2er} \left[ \tau^k_\phi 
 \left( H_1 dr + \left(1-H_2\right) r d\theta \right)
 -k \left( \tau^k_r H_3 + \tau^k_\theta \left(1-H_4\right) \right)
  r \sin \theta d\phi \right]
\ , \label{gf1} \end{equation}
with the Pauli matrices $\vec \tau = ( \tau_x, \tau_y, \tau_z) $ and
$\tau^k_r = \vec \tau \cdot 
(\sin \theta \cos k \phi, \sin \theta \sin k \phi, \cos \theta)$,
$\tau^k_\theta = \vec \tau \cdot 
(\cos \theta \cos k \phi, \cos \theta \sin k \phi, -\sin \theta)$,
$\tau^k_\phi = \vec \tau \cdot (-\sin k \phi, \cos k \phi,0)$.
We refer to $k$ as the winding number of the solutions.
Again, the four gauge field functions $H_i$ 
and the dilaton function $\Phi$ depend only on $r$ and $\theta$.
For $k=1$ the spherically symmetric ansatz of ref.~\cite{eymd} 
is recovered with $H_1=H_3=0$, $H_2=H_4=w(r)$ and $\Phi=\Phi(r)$.

Denoting the stress-energy tensor of the matter fields by $T_{\mu}^{\nu}$,
with this ansatz the energy density $\epsilon =-T_0^0=-L_M$ becomes
\begin{eqnarray}
-T_0^0 = & \frac{f}{2m} \left[
 (\partial_r \Phi )^2 + \frac{1}{r^2} (\partial_\theta \Phi )^2 \right]
       + e^{2 \kappa \Phi} \frac{f^2}{2 e^2 r^4 m} \left\{
 \frac{1}{m} \left(r \partial_r H_2 + \partial_\theta H_1\right)^2 
 \right.
\nonumber \\
      & +  \left.
   \frac{k^2}{l} \left [
  \left(  r \partial_r H_3 - H_1 H_4 \right)^2
+ \left(r \partial_r H_4 + H_1 \left( H_3 + {\rm ctg} \theta \right)
    \right)^2 \right. \right.
\nonumber \\
      & +  \left. \left.
  \left(\partial_\theta H_3 - 1 + {\rm ctg} \theta H_3 + H_2 H_4
     \right)^2 +
  \left(\partial_\theta H_4 + {\rm ctg} \theta \left( H_4-H_2 \right) 
   - H_2 H_3 \right)^2 \right] \right\}
\ . \label{edens} \end{eqnarray}

With respect to 
$ U= e^{i\Gamma(r,\theta) \tau^k_\phi} $
the system possesses a residual abelian gauge invariance.
To fix the gauge we choose the
gauge condition 
$ r \partial_r H_1 - \partial_\theta H_2 = 0 $.

\subsection{Regular solutions}

We first briefly consider static axially symmetric solutions
of the field equations, which have a finite mass 
and are globally regular and asymptotically flat.

We change to dimensionless quantities again.
The dimensionless mass $\mu =(e/\sqrt{4\pi G}) G M$
is determined by the derivative of the metric function $f$
at infinity,
$\mu = \frac{1}{2} x^2 \partial_x f |_\infty $.
Similarly, the derivative of the dilaton function at infinity
determines the dilaton charge $D  =  x^2 \partial_x \varphi |_\infty $.
 
As for the spherically symmetric solutions, the following relations 
between the metric and the dilaton field hold \cite{kks2}
\begin{equation}
\varphi(x) = \frac{1}{2} \gamma \ln(-g_{tt})
\ , \label{res2} \end{equation}
\begin{equation}
D = \gamma \mu 
\ . \label{res1} \end{equation}

For the globally regular EYM solution with $k=2$ and $n=1$ 
the energy density of the matter fields is shown in Fig.~4,
together with a surface of constant energy density,
which is toruslike.
\begin{figure}[t]
\vspace*{-4.5cm}
\hspace*{-2cm}
\centering
\epsfysize=4in
\mbox{\epsffile{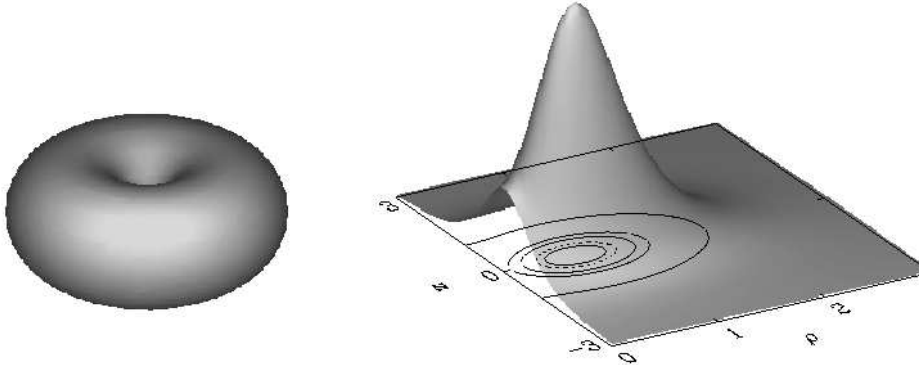}}
\caption{
A surface of constant energy density (left)
and the energy density (right) of the globally regular EYM solution
with $k=2$ and $n=1$.
}
\end{figure}

\subsection{Black hole solutions}

Now we consider static axially symmetric black hole solutions
with a regular event horizon.
The event horizon of static black hole solutions
is characterized by $g_{tt}=-f=0$.
We impose that the regular horizon resides at a surface of constant $r$.

According to the zeroth law of black hole physics,
the surface gravity $\kappa_{\rm sg}$ \cite{wald,ewein},
\begin{equation}
\kappa^2_{\rm sg}=-(1/4)g^{tt}g^{ij}(\partial_i g_{tt})(\partial_j g_{tt})
\ , \label{sg} \end{equation}
is constant at the horizon of the black hole solutions.
It is proportional to the temperature $T=\kappa_{\rm sg} /(2 \pi)$.
The dimensionless area $A$ 
of the event horizon of the black hole solutions
is proportional to the entropy $S = A/4$ \cite{wald}.
The black hole solutions satisfy the general mass relation
\begin{equation}
\mu  =\mu_{\rm o} + 2 TS
\ , \label{mass2} \end{equation}
with $\mu_{\rm o}= - (e/\sqrt{4 \pi G}) G 
  \int_0^{2 \pi} \int_0^\pi \int_{r_{\rm H}}^\infty
  d \phi d \theta d r
  \sqrt{-g} \left( 2 T_0^{\ 0} - T_\mu^{\ \mu} \right)$
\cite{wald},
and the relation (compare (\ref{res1}))
\begin{equation}
D =  \gamma ( \mu - 2 TS ) 
\ . \end{equation}

In Fig.~5 we exhibit surfaces of constant energy density
and in Fig.~6 the energy density 
for the EYM black hole solution with $k=2$ and $n=1$.
Also the event horizon of the black hole solutions
is not spherically symmetric, but only axially symmetric \cite{kk3}.
\begin{figure}[t]
\vspace*{-1cm}
\centering
\epsfysize=4.5in
\mbox{\epsffile{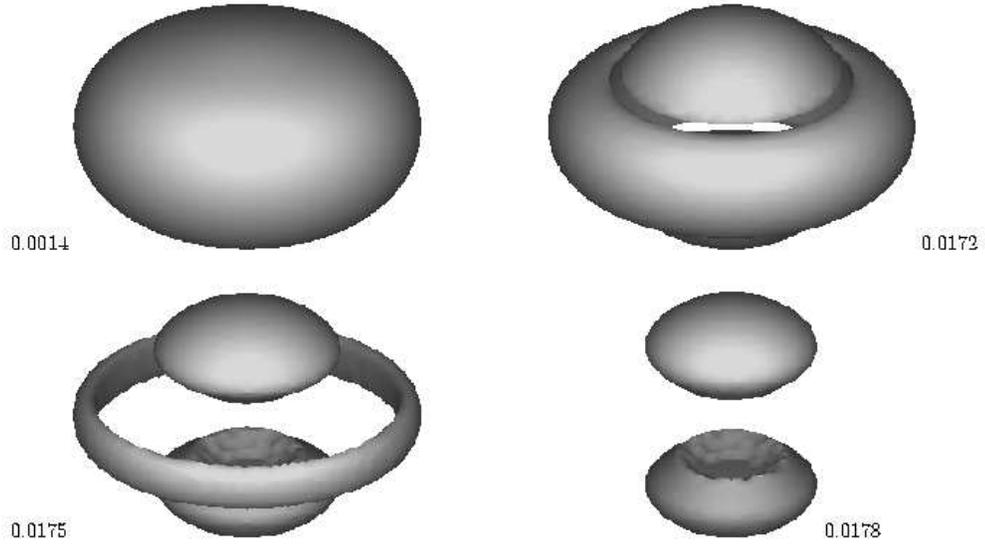}}
\vspace*{-2cm}
\caption{
Surfaces of constant energy density 
for four values of the dimensionless energy density
of the EYM black hole solution
with $k=2$ and $n=1$.
}
\end{figure}
\begin{figure}[h!]
\vspace*{-0.5cm}
\centering
\epsfysize=5in
\mbox{\epsffile{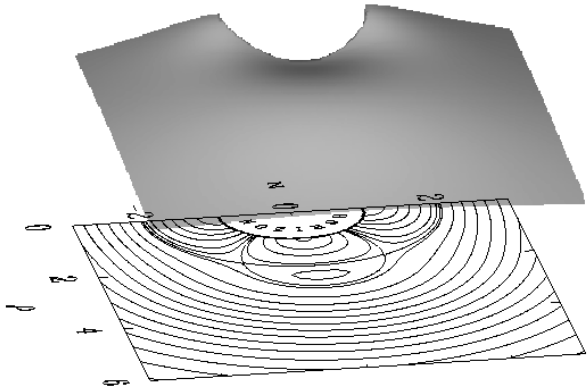}}
\vspace*{-5cm}
\caption{
The energy density of the EYM black hole solution
with $k=2$ and $n=1$.
}
\end{figure}

\section{Conclusions}

In EYM and EYMD theory black hole solutions are not uniquely
specified by their mass, charge and angular momentum.
Here we have considered static EYM and EYMD black hole solutions
with non-abelian hair.
The spherically symmetric solutions are unstable \cite{strau},
and there is all reason to believe, that the axially symmetric
black hole solutions are unstable as well.
However, we expect analogous solutions in EYMH theory \cite{ewein}
and in ES theory,
and for $n=2$ these axially symmetric black hole solutions
should be stable \cite{ewein}.
In contrast, the stable black hole solutions
with higher magnetic charges (EYMH) or higher baryon numbers (ES)
should not correspond to such axially symmetric solutions with $n>2$.
Instead these stable solutions should have much more complex shapes,
exhibiting discrete crystal-like symmetries \cite{ewein}.
Analogous but unstable black hole solutions of this kind
should also exist in EYM and EYMD theory.

\end{document}